\documentclass[12pt]{article}
\usepackage{amssymb}
\usepackage[dvips]{epsfig}

\newcommand{\be}{\begin{equation}}
\newcommand{\ee}{\end{equation}}
\newcommand{\bn}{\begin{eqnarray}}
\newcommand{\en}{\end{eqnarray}}

\newcommand{\p}{\partial}

\newcommand{\csl}{CS$_1$ }

\newcommand{\tj}{\tilde{j} }
\newcommand{\fab}{f_{\alpha\beta} }

\newcommand{\nn}{\nonumber}

\newcommand{\no}{\noindent}

\def\bea{\begin{eqnarray}}
\def\eea{\end{eqnarray}}

\newcommand{\beq}{\begin{eqnarray}}
\newcommand{\eeq}{\end{eqnarray}}
\begin{document}

\title{\textbf{Dual descriptions of  spin two massive particles in
$D=2+1$ via master actions}}
\author{D. Dalmazi  and Elias L. Mendon\c ca \\
\textit{{UNESP - Campus de Guaratinguet\'a - DFQ} }\\
\textit{{Av. Dr. Ariberto Pereira da Cunha, 333} }\\
\textit{{CEP 12516-410 - Guaratinguet\'a - SP - Brazil.} }\\
\textsf{E-mail: dalmazi@feg.unesp.br , elias.fis@gmail.com }}
\date{\today}
\maketitle

\begin{abstract}

In the first part of this work we show the decoupling (up to
contact terms) of redundant degrees of freedom which appear in the
covariant description of spin two massive particles in $D=2+1$. We
make use of a master action which interpolates, without solving
any constraints, between a first, second and third order (in
derivatives) self-dual model. An explicit dual map between those
models is derived. In our approach the absence of ghosts in the
third order self-dual model, which corresponds to a quadratic
truncation of topologically massive gravity, is due to the
triviality (no particle content) of the Einstein-Hilbert action in
$D=2+1$. In the second part of the work, also in $D=2+1$, we prove
the quantum equivalence of the gauge invariant sector of a couple
of self-dual models of opposite helicities ($+2$ and $-2$) and
masses $m_+$ and $m_-$ to a generalized self-dual model which
contains a quadratic Einstein-Hilbert action, a Chern-Simons term
of first order and a Fierz-Pauli mass term. The use of a first
order Chern-Simons term instead of a third order one avoids
conflicts with the sign of the Einstein-Hilbert action.

\end{abstract}

\newpage

\section{Introduction}

In the last years there has been a quite intense activity in the
subject of higher spin theories  in different dimensions and their
dual formulations, see for instance
\cite{vasiliev,sandrine,zinoviev,gaitan,montemayor} and references
therein. One of the difficulties of a covariant description of
higher spin fields is the amount of redundant degrees of freedom
present in the higher rank tensor fields. This is a severe
difficulty in constructing interacting theories for such fields,
see comments in \cite{zinoviev,gaitan}. In the first part of our
work (section 2) we address the issue of spurious degrees of
freedom in $D=2+1$ for massive fields of helicity $\pm 2$. We show
how duality can help us to prove the quantum decoupling of
redundant degrees of freedom at quadratic level (free theories).
Our master action approach leads us also to a better understanding
of the differences with the spin one case where there are only
first order and second order (in derivatives) self-dual models
unlike the spin two case where we have also a third order (ghost
free) self-dual model. In particular, based on the local
symmetries of the dual models we also argue why we do not expect a
fourth (or higher) order self-dual model for spin two and why we
do not have a third (or higher) order  self-dual model for the
spin one case. Our approach makes clear that the absence of ghosts
in the third order self-dual model is a consequence of the
non-propagating nature of the Einstein-Hilbert action in $D=2+1$.

In the second part of this work (section 3) we show that there
exists a self-consistent quantum description of a couple of
massive states of opposite helicities ($+2$ and $-2$) and
different masses in general, by means of only one rank two tensor
field which we call a generalized self-dual (GSD) field in analogy
with the spin one case treated in \cite{kumar,jhep2}. We avoid the
conflicts found in  \cite{tekin} with the sign of the
Einstein-Hilbert term by working with a Chern-Simons term of first
order instead of the gravitational Chern-Simons term of third
order of \cite{djt}. The particle content of the GSD model is
disentangled by showing
 its dual equivalence to
the gauge invariant sector of a couple of non-interacting second
order self-dual models of opposite helicities.

\section{First, second and third order self-dual models and their dual maps}

 Our starting point is the first order self-dual model
suggested in \cite{aragone} which is the helicity $+ 2$ analogue
of the helicity $+1$ self-dual model of \cite{tpn},

\be S_{SD}^{(1)} = \int  d^3 x \left\lbrack \frac m2
\epsilon^{\mu\nu\lambda}
f_{\mu}\,^{\alpha}\p_{\nu}f_{\lambda\alpha} + \frac{m^2}2
\left(f^2 - f_{\mu\nu}f^{\nu\mu}\right) \right\rbrack \qquad ,
\label{lsd1} \ee

\no where $f\equiv \eta^{\alpha\beta} f_{\alpha\beta}$. The metric
is flat: $\eta_{\alpha\beta}={\rm diag}\left(-,+,+\right)$. The
upper index in $S_{SD}^{(1)}$ indicates that we have a first order
model in the derivatives. In most of this work we use second rank
tensor fields, like $\fab$ in (\ref{lsd1}), with no symmetry in
their indices. Whenever symmetric and antisymmetric combinations
show up they will be denoted respectively by:
$f_{(\alpha\beta)}\equiv \left( f_{\alpha\beta} + f_{\beta\alpha}
\right)/2 $ and $f_{\lbrack \alpha\beta\rbrack}\equiv \left(
f_{\alpha\beta} - f_{\beta\alpha} \right)/2 $. Replacing $m$ by
$-m$ in $S_{SD}^{(1)}$ we change the particle's helicity from $+2$
to $-2$. The first term in (\ref{lsd1}) reminds us of a spin one
topological Chern-Simons term which will be called henceforth a
Chern-Simons term of first order (CS$_1$), to be distinguished
from another (third order) Chern-Simons term which appears later.
The second term in (\ref{lsd1}) is the Fierz-Pauli (FP) mass term
\cite{fierz} which is the spin two analogue of a spin one Proca
mass term. The FP term breaks the local invariance $\delta
f_{\alpha\beta} = \p_{\alpha}\xi_{\beta} $ of the \csl term.

The equations of motion of (\ref{lsd1}),

\be \epsilon_{\mu}\,^{\nu\lambda} \p_{\nu}f_{\lambda\alpha} =
m\left( f_{\alpha\mu} - \eta_{\mu\alpha}\, f \right) \qquad ,
\label{eq1} \ee

\no  imply that $f_{\alpha\beta}$ is traceless, symmetric and
transverse, i.e.,

 \bea
f &=& 0 \label{trace}\\
f_{\lbrack \alpha\beta\rbrack} &=& 0 \label{anti}\\
\p^{\alpha}f_{\alpha\beta} &=& 0 \,= \,\p^{\beta}f_{\alpha\beta}
\label{transverse}\eea

\no Furthermore, it follows that $f_{\alpha\beta}$ satisfies the
Klein-Gordon equation $\left(\Box - m^2\right) f_{\alpha\beta}= 0$
and the helicity equation $\left( J^{\mu}P_{\mu} + 2 \, m
\right)^{\alpha\beta\gamma\delta} f_{\gamma\delta}=0$, with
$(2m)^{\alpha\beta\gamma\delta} = m \left( \delta^{\alpha\gamma}
\delta^{\beta\delta} + \delta^{\alpha\delta}
\delta^{\beta\gamma}\right)$ and, see \cite{gaitan}, the
quantities $\left(J^{\mu}\right)^{\alpha\beta\gamma\delta} = i\,
\left(\eta^{\alpha\gamma}\epsilon^{\beta\mu\delta} +
\eta^{\beta\gamma}\epsilon^{\alpha\mu\delta} +
\eta^{\alpha\delta}\epsilon^{\beta\mu\gamma} +
\eta^{\beta\delta}\epsilon^{\alpha\mu\gamma} \right)/2$ satisfy
the $2+1$ Lorentz algebra.
 In summary, all necessary equations to
describe a helicity $+2$  massive particle in $D=2+1$ are
satisfied at classical level.

Next we combine the works \cite{aragone} and \cite{deser} into one
master action which takes us from the first order self-dual model
(\ref{lsd1}) to its second and third order version  entirely
within the path integral framework with no need of solving any
constraint equation as in \cite{aragone} or introducing any
explicit gauge condition. Before we proceed, in order to keep the
analogy with the spin one case as close as possible and to avoid
the profusion of indices we use the shorthand notation:

\bea \int f \cdot d\, f \, &\equiv & \int  d^3 x \,
\epsilon_{\mu}\,^{\nu\lambda}
f^{\mu\alpha}\p_{\nu}f_{\lambda\alpha}
\label{csl1}\\
\int \left(f^2 \right)_{FP} \, &\equiv & \int  d^3 x \, \left(f^2
- f_{\mu\nu}f^{\nu\mu}\right) \label{fp} \eea

\no In the master action approach an important role will be played
by the Einstein-Hilbert (EH) term. If we expand in the dreibein
$e_{\mu\alpha} = \eta_{\mu\alpha} + h_{\mu\alpha}$ and keep only
quadratic terms in the fluctuations, the EH action can be written
\cite{deser}:

\be - \frac 12 \int \, d^3 x \, \left( \sqrt{- g} \, R
\right)_{hh} = \int \, d^3 x \frac {\epsilon^{\mu\nu\lambda}
h_{\mu}\, ^{\alpha}\p_{\nu} \Omega_{\lambda\alpha}(h)}{4} = \frac
14 \int h \cdot d\, \Omega (h) \qquad , \label{leh} \ee

\no where

\be \Omega_{\lambda}\, ^{\alpha}(h) =
\epsilon^{\alpha\beta\gamma}\left\lbrack
\p_{\lambda}h_{\gamma\beta} - \p_{\beta}\left(h_{\gamma\lambda} +
h_{\lambda\gamma}\right)\right\rbrack \label{omega}\ee

\no As explained in \cite{jhep1,jhep2} with an explicit example,
the existence of a master action does not guarantee {\it a priori}
spectrum equivalence of the interpolated dual theories. It is
crucial that the terms which mix the fields of the dual theories
have no propagating degree of freedom like the spin one CS term
used in \cite{dj} or the BF type mixing terms of \cite{botta}.
Based on the works \cite{aragone} and \cite{deser} we suggest the
following master action:

\bea S_{M}^S &=&  \frac m2 \int f \cdot d\, f
+ \frac{m^2}2 \int \left(f^2 \right)_{FP} - \frac m2  \int (f-A) \cdot d\, (f-A)\nn \\
&-&   a \, \int \left(h-A\right)\cdot d\, \Omega (h-A) \label{sm1}
\qquad . \eea

\no We have introduced two second rank tensor fields
$A_{\alpha\beta}$ and $h_{\alpha\beta}$ with no symmetry in their
indices. The upper index in $S_M^S$ stands for singlet (parity
singlet of helicity +2). The coefficient in front of the third
term of (\ref{sm1}) is such that the quadratic term of $S_M^S$ in
$f_{\alpha\beta}$ has no derivatives which is important for
deriving dual theories which are local. The constant ``$a$'' will
be fixed later on for an analogous reason. If $a=0$ we recover the
intermediate master action of \cite{aragone}.  Let us introduce
sources $j_{\alpha\beta}$ and define the generating function:

\be W^S \left\lbrack J \right\rbrack = \int {\cal
D}A_{\alpha\beta} \, {\cal D}h_{\alpha\beta}\,  {\cal
D}f_{\alpha\beta} \, \exp i \left( S_M^S + \int\, d^3 x
f_{\alpha\beta}\, j^{\alpha\beta} \right) \label{ws1} \ee

\no After the trivial shift $h_{\alpha\beta} \to h_{\alpha\beta} +
A_{\alpha\beta}$ followed by $A_{\alpha\beta} \to A_{\alpha\beta}
+ f_{\alpha\beta}$, the last two terms of (\ref{sm1}) decouple and
since they have no particle content it is clear that $S_M^S $ is
equivalent to $S_{SD}^{(1)}$ and therefore describes a parity
singlet of helicity $+2$. After those shifts and integrating over
$h_{\alpha\beta}$ and $A_{\alpha\beta}$ we derive, up to an
overall constant,

\be W^S \left\lbrack J \right\rbrack = \int \,  {\cal
D}f_{\alpha\beta} \exp i \left( S_{SD}^{(1)} + \int\, d^3 x
f_{\alpha\beta}j^{\alpha\beta} \right) \qquad . \label{ws2} \ee

\no On the other hand, since the linear term in the fields $\fab$
in the exponent in (\ref{ws1}) is $\fab U^{\alpha\beta}$ with
$U^{\alpha\beta} \equiv m \, \epsilon^{\alpha\nu\lambda}\p_{\nu}
A_{\lambda}\,^{\beta} + j^{\alpha\beta}$, after the shift $\fab
\to \fab + \left(\eta_{\alpha\beta} U_{\mu}^{\mu} - 2 \,
U_{\alpha\beta} \right)/(2m^2)$  we decouple $\fab$ completely.
After integrating over $\fab$ we obtain, up to an overall
constant,

\be W^S \left\lbrack J \right\rbrack = \int \, {\cal
D}A_{\alpha\beta} \, {\cal D}h_{\alpha\beta} \, \exp i \, S_I
\left\lbrack j \right\rbrack \qquad , \label{wsi} \ee

\no where

\bea S_I \left\lbrack j \right\rbrack &=& \int \left\lbrack
\frac{A \cdot d\, \Omega (A)}4 - \frac m2 A \cdot d\, A
\right\rbrack - a \, \int \left(h-A\right)\cdot d\, \Omega
(h-A)\nn\\
&+& \int \, d^3 x \, \left\lbrack j^{\alpha\beta}
F_{\alpha\beta}(A) + \frac{j^{\alpha\beta} j_{\beta\alpha}}{2 m^2}
- \frac{ (j_{\mu}^{\mu})^2}{4 m^2} \right\rbrack \label{si} \eea

\no The sources are now coupled to the gauge invariant
combination:

\be F_{\alpha\beta}(A) \equiv T_{\alpha\beta}(A)-
\frac{T_{\mu}\,^{\mu}(A)}{2} \eta_{\alpha\beta} \label{fab} \ee

\no where $T_{\beta\alpha}(A) \equiv (\frac 1m)
\epsilon_{\beta}\,^{\nu\lambda}\p_{\nu} A_{\lambda\alpha}$ is
invariant under the gauge transformations $\delta A_{\alpha\beta}
= \p_{\alpha}\xi_{\beta}$. The shift $h_{\alpha\beta} \to
h_{\alpha\beta} + A_{\alpha\beta} $ in (\ref{si}) decouples
$h_{\alpha\beta}$ for arbitrary values of the constant ``$a$'',
which has played no role so far. Integrating $h_{\alpha\beta}$, up
to an overall constant again, we obtain

\be W^S \left\lbrack J \right\rbrack = \int \, {\cal
D}A_{\alpha\beta} \exp i \left\lbrace S_{SD}^{(2)}(A) + \int \,
d^3 x \, \left\lbrack j^{\alpha\beta} F_{\alpha\beta}(A) +
\frac{j^{\alpha\beta} j_{\beta\alpha}}{2 m^2} - \frac{
(j_{\mu}^{\mu})^2}{4 m^2} \right\rbrack \right\rbrace \qquad ,
\label{ws3} \ee

\no  where the second order self-dual model is given by:

\be S_{SD}^{(2)} = \int \left\lbrack \frac{A \cdot d\, \Omega
(A)}4 - \frac m2 A \cdot d\, A \right\rbrack \qquad .
\label{sd2}\ee

\no The model $S_{SD}^{(2)}$ has appeared before in
\cite{aragone,deser}. It looks very similar to the spin one
Maxwell-Chern-Simons (MCS) theory of \cite{djt}. In particular,
$S_{SD}^{(2)}$ is a gauge theory invariant under  $\delta
A_{\alpha\beta} = \p_{\alpha}\xi_{\beta} $. The first term in
(\ref{sd2}) is the analogue of the Maxwell term in the MCS theory
and corresponds exactly to the quadratic approximation of the
Einstein-Hilbert action, see (\ref{leh}), with its usual sign.

 From the classical point of view, the equations of motion of
$S_{SD}^{(2)}$ can be cast in the same self-dual form (\ref{eq1})
with the identification $\fab \leftrightarrow F_{\alpha\beta}(A)$.
Therefore, it is clear that $S_{SD}^{(2)}$ is a perfectly
acceptable classical description of such particle. It is amazing
that although each of the terms in (\ref{sd2}) has no particle
content, when they are put together they describe a massive
particle of helicity $+2$.

At quantum level, deriving (\ref{ws2}) and (\ref{ws3}) with
respect to the sources we demonstrate the following equivalence of
correlation functions:

\be \left\langle  f_{\mu_1\nu_1}(x_1)\cdots
f_{\mu_N\nu_N}(x_N)\right\rangle_{S_{SD}^{(1)}} = \left\langle
F_{\mu_1\nu_1}\left\lbrack A(x_1)\right\rbrack \cdots
F_{\mu_N\nu_N}\left\lbrack
A(x_N)\right\rbrack\right\rangle_{S_{SD}^{(2)}}  \, + \, {\rm
contact \, \, terms} \label{cf1}\ee

\no The contact terms appear due to the quadratic terms in the
sources in (\ref{ws3}). In conclusion, we have the dual map below
at classical and quantum level,

\be \fab \leftrightarrow F_{\alpha\beta}(A) = T_{\alpha\beta}(A)-
\frac{T_{\mu}^{\mu}(A)}{2} \eta_{\alpha\beta} \qquad .
\label{dualmap1} \ee

\no Due to the gauge invariance of $T_{\alpha\beta}(A) =
\epsilon_{\alpha}\, ^{\nu\lambda}\p_{\nu} A_{\lambda\beta}/m $ our
dual map is gauge invariant as expected since $S_{SD}^{(1)}$ is
not a gauge theory. The map (\ref{dualmap1}) is similar to the
spin one map
$f_{\mu}\leftrightarrow\epsilon_{\mu\nu\alpha}\p^{\nu}A^{\alpha}/m$
between the self-dual model of \cite{tpn} and the MCS theory of
\cite{djt}.

Next we show that $S_{SD}^{(1)}$ is also dual to a third order
self-dual model. Neglecting surface terms, after some integration
by parts it is easy to prove the identities:

\be \int h \cdot d\, \Omega (A) = \int A \cdot d\, \Omega (h) =
\int \Omega (h)  \cdot d\, A \label{identities} \ee

\no By using those identities in (\ref{si}) and fixing $a=1/4$, we
can cancel the second order term $\int A \cdot d \, \Omega (A)/4$
and the intermediate action (\ref{si}) can be written as:

\bea S_I \left\lbrack j \right\rbrack &=& -\frac m2 \int
\left\lbrack A - \frac{\Omega (h)}{2 m} \right\rbrack \cdot d\,
\left\lbrack A - \frac{\Omega (h)}{2 m} \right\rbrack \nn \\
&+& \frac 1{8m} \int \Omega (h) \cdot d \,   \Omega (h)  - \frac
14 \int h \cdot d \,   \Omega (h) \nn \\
&+& \int \, d^3 x \, \left\lbrack j^{\alpha\beta}
F_{\alpha\beta}(A) + \frac{j^{\alpha\beta} j_{\beta\alpha}}{2 m^2}
- \frac{ (j_{\mu}^{\mu})^2}{4 m^2} \right\rbrack \label{si2} \eea

\no It is clear that the shift $A_{\alpha\beta} \to
A_{\alpha\beta} + \Omega_{\alpha\beta}(h)/2m $ will decouple
$A_{\alpha\beta}$ from  $h_{\alpha\beta} $ and produces the third
order action $\int \Omega d \Omega$ out of the second order theory
(\ref{si}). Another, less obvious, shift $A_{\alpha\beta} \to
A_{\alpha\beta} + \left( j_{\beta\alpha} -
\eta_{\beta\alpha}j_{\mu}^{\mu}/2\right)/m^2 $ decouples
$A_{\alpha\beta}$ completely and gives rise to the CS$_1$
 term $-(m/2)\int A \cdot d \, A$ with no particle content. After
integrating over $A_{\alpha\beta}$ we derive from (\ref{wsi}) and
(\ref{si2}), up to an overall constant,

\be W^S \left\lbrack J \right\rbrack = \int \, {\cal
D}h_{\alpha\beta} \exp i \left\lbrack S_{SD}^{(3)}(h) + \int \,
d^3 x \, \left\lbrack j^{\alpha\beta}
F_{\alpha\beta}\left(\frac{\Omega }{2 m}\right) + {\cal O}(j^2)
\right\rbrack \right\rbrace \qquad , \label{ws4} \ee

\no where ${\cal O}(j^2)$ stands for quadratic terms in the
sources which lead only to contact terms in the correlation
functions and therefore do not need to be specified. From
(\ref{fab}) and (\ref{omega}) we have:

\bea F_{\alpha\beta}\left( \frac{\Omega }{2m}\right) &=&
T_{\alpha\beta}\left(\frac{\Omega}{2m}\right)-
\frac{T_{\mu}\,^{\mu}\left(\frac{\Omega }{2m}\right)}{2} \eta_{\alpha\beta} \qquad , \label{fabomega}\\
T^{\alpha\beta}\left(\frac{\Omega }{2m}\right) &=& \frac
{\epsilon^{\alpha\nu\lambda}\p_{\nu} \Omega_{\lambda}\,^{\beta}}{2
m^2} = - \frac{E^{\alpha\gamma}E^{\beta\lambda}
h_{(\gamma\lambda)}}{m^2} \qquad , \label{tabomega} \eea

\no with $E^{\lambda\mu} \equiv \epsilon^{\lambda\mu\nu}\p_{\nu}$.
The third order self-dual model $S_{SD}^{(3)}(h)$ is given by:

\bea S_{SD}^{(3)}(h) &=& \frac 1{8m} \int \Omega (h) \cdot d \,
\Omega (h)  - \frac 14 \int h \cdot d \,   \Omega (h)
\nn \\
&=&  \int d^3 x \left\lbrack -\frac 1{4m} h_{(\lambda\mu)}
\left(\eta^{\lambda\delta} \Box - \p^{\lambda}\p^{\delta} \right)
E^{\mu\alpha} h_{(\alpha\delta)} + \frac 12 h_{(\lambda\mu)}
E^{\lambda\delta} E^{\mu\alpha} h_{(\alpha\delta)}
\right\rbrack\label{sd3} \eea

\no  The first term in $S_{SD}^{(3)}(h)$ is the quadratic
approximation in the fluctuations of the dreibein $e_{\alpha\beta}
= \eta_{\alpha\beta} + h_{\alpha\beta}$ of a gravitational
Chern-Simons term, see \cite{djt,deser}, while the second one is
the Einstein-Hilbert (EH) term at the same approximation, see
(\ref{leh}). Both terms form the quadratic approximation for the
so called topologically massive gravity (TMG) of \cite{djt}. The
action $S_{SD}^{(3)}$ is invariant under the local transformations
$\delta h_{\alpha\beta} = \p_{\alpha}\xi_{\beta} +
\epsilon_{\alpha\beta\gamma}\phi^{\gamma}$. Notice that the sign
of the EH term is not the expected one. By construction, in
passing from $S_{SD}^{(2)}(h)$ to $S_{SD}^{(3)}(h)$ there is a
sign inversion. The unexpected sign, as explained in \cite{djt},
is in fact necessary for absence of ghosts which is a surprising
feature of the higher order theory $S_{SD}^{(3)}(h)$ that we now
understand from another point of view, since we have shown
directly that $S_{SD}^{(3)}(h)$ can be derived from the first
order ghost free theory $S_{SD}^{(1)}(h)$ by the addition of two
extra terms (mixing terms), see (\ref{sm1}), with no particle
content. Now it is clear why we do not have a third order
self-dual model in the spin one case, the point is that when we
derive $S_{SD}^{(3)}(h)$ from a first order theory a second order
mixing term is necessary. We have used the quadratic
Einstein-Hilbert action as a mixing term since it has no particle
content. However, its spin one analogue is the Maxwell action
which contains a scalar massless particle in the spectrum and can
not be used to mix dual fields without leading to a spectrum
mismatch between the dual theories.

At classical level, the equations of motion $\delta
S_{SD}^{(3)}=0$ can be written in the first order self-dual form
(\ref{eq1}) with the identification $\fab \leftrightarrow
F_{\alpha\beta}\left( \frac{\Omega }{2m} \right)$. Consequently,
$S_{SD}^{(3)}$ describes classically a parity singlet of helicity
$+2$ just like $S_{SD}^{(2)}$ or $S_{SD}^{(1)}$.

From  (\ref{ws2}) and (\ref{ws4}) we deduce:

\be \left\langle  f_{\mu_1\nu_1}(x_1)\cdots
f_{\mu_N\nu_N}(x_N)\right\rangle_{S_{SD}^{(1)}} = \left\langle
F_{\mu_1\nu_1}\left\lbrack \frac{\Omega(x_1)}{2m}\right\rbrack
\cdots F_{\mu_N\nu_N}\left\lbrack
\frac{\Omega(x_N)}{2m}\right\rbrack\right\rangle_{S_{SD}^{(3)}} \,
+ \, {\rm contact \, \, terms} \label{cf2}\ee

\no It is remarkable that now in the $S_{SD}^{(3)}(h)$ theory we
have $T_{\alpha\beta}\left(\frac{\Omega }{2m} \right)
=T_{\beta\alpha}\left( \frac{\Omega }{2m} \right)$, see
(\ref{tabomega}), and consequently $F_{\alpha\beta}\left(
\frac{\Omega }{2m} \right) =F_{\beta\alpha}\left( \frac{\Omega
}{2m} \right)$. Therefore the dual map $f_{\alpha\beta}
\leftrightarrow F_{\alpha\beta}\left( \frac{\Omega}{2m}\right)$
that we read from (\ref{cf2}) now automatically assures the
vanishing  of correlation functions of the antisymmetric
combinations $f_{\lbrack \alpha\beta\rbrack}$, up to contact
terms, which is not obvious neither in $S_{SD}^{(1)}(f)$ nor in
$S_{SD}^{(2)}(A)$. This is a typical advantage of having dual
formulations of the same theory.

The decoupling of the trace $f=\eta^{\alpha\beta}\fab$ is not
obvious in none of the three self-dual formulations given here. In
what follows we take advantage of the second order formulation to
prove it. First, suppose we had defined the sources from the very
beginning as $j^{\alpha\beta} \equiv \phi \eta^{\alpha\beta} +
j^{\alpha\beta}_S + j^{\alpha\beta}_A $, such that $\fab
j^{\alpha\beta} = f\, \phi + j^{\alpha\beta}_S f_{(\alpha\beta)} +
j^{\alpha\beta}_A f_{\lbrack \alpha\beta\rbrack} $ where
$j^{\alpha\beta}_S=j^{\beta\alpha}_S $ and $j^{\alpha\beta}_A = -
j^{\beta\alpha}_A$. Back in (\ref{ws3}) and using (\ref{fab}) we
can write down the action in the exponent of (\ref{ws3}) as
follows:

\bea S\left\lbrack j \right\rbrack &=& \int \, d^3 x \left\lbrack
- \frac{A_{\mu\alpha}
E^{\mu\lambda}E^{\alpha\gamma}\left(A_{\gamma\lambda} +
A_{\lambda\gamma}\right)}{4} -
m^2\frac{A_{\mu\alpha}T^{\mu\alpha}(A)}2
 \right.  \nn\\ &+& j_A^{\mu\alpha}T_{\mu\alpha}(A) + j_S^{\mu\alpha}T_{\mu\alpha}(A)
  - \left. \frac{\lbrack \phi + (j_S)_{\nu}^{\nu} \rbrack T_{\mu}^{\mu}}2 +
{\cal O}(j_{\alpha\beta}^2)  \right\rbrack \label{sj} \eea

\no Since the first term in (\ref{sj}), which is the quadratic
Einstein-Hilbert action, only depends on $A_{(\mu\alpha)}$ it is
clear that we get rid of $j_A^{\mu\alpha}T_{\mu\alpha}(A)$ through
the shift $A^{\mu\alpha} \to A^{\mu\alpha} + j^{\mu\alpha}_A/m^2$.
So we can see the decoupling of $f_{\lbrack \alpha\beta\rbrack}$
directly in the $S_{SD}^{(2)}$ formulation. After $A^{\mu\alpha}
\to A^{\mu\alpha} + \left( \frac{E^{\mu\alpha}}{m} -
\eta^{\mu\alpha} \right) \frac{\phi}{2 m^2}$ we cancel out
 $-\phi \, T_{\mu}^{\mu}/2$ in (\ref{sj}). Consequently, all correlation functions
  of $f_{\lbrack \alpha\beta\rbrack}$ or the trace $f$ will vanish, up to contact terms, in agreement
   with the classical results (\ref{trace}) and (\ref{anti}).

  Regarding the transverse condition (\ref{transverse}), from the trace of the dual map (\ref{dualmap1})
   we have the correspondence $f \leftrightarrow
  -T_{\mu}^{\mu}(A)/2$. So, the decoupling of the trace $f$ implies that correlation functions in
  the $S_{SD}^{(2)}(A)$ theory involving
  $T_{\mu}^{\mu}(A)$ must vanish (up to contact terms). Classically, $T_{\mu}^{\mu}(A)=0$ follows from the
  equations of motion of $S_{SD}^{(2)}(A)$. Thus, we can reduce the dual map
  (\ref{dualmap1}) to $\fab \leftrightarrow T_{\alpha\beta}(A)$.
  Due to the trivial (non-dynamical) identity $\p_{\alpha}
  T^{\alpha\beta}=0$ it follows $\p_{\alpha}
  f^{\alpha\beta}=0$ and since $f^{\lbrack \alpha\beta\rbrack}$
  decouples we have $\p_{\alpha}
  f^{\alpha\beta}=0=\p_{\alpha}
  f^{\beta\alpha}=0$ inside correlation functions up to contact
  terms. Therefore all constraints
  (\ref{trace}),(\ref{anti}) and (\ref{transverse}) are satisfied. We can use the dual maps between correlation
  functions (\ref{cf1}) and (\ref{cf2})
  and the detailed studies (including the pole structure of the propagator) made in
\cite{djt}, see also \cite{aragone}, to finally establish that the
three models $S_{SD}^{(1)}(f)$, $\, S_{SD}^{(2)}(A)$ and
$S_{SD}^{(3)}(h)$  correctly describe a parity singlet of helicity
$+2$ and mass $m$.

The fact that (\ref{anti}) and (\ref{transverse}) are consequences
of trivial (non-dynamical) identities is relevant for a consistent
coupling to other fields. In the spin one case the transverse
condition on the self-dual field $\p_{\mu}f^{\mu}=0$ is traded, in
the Maxwell-Chern-Simons theory, in the Bianchi identity
$\p_{\mu}F^{\mu}(A)=\p_{\mu}\left(\epsilon^{\mu\nu\alpha}\p_{\nu}A_{\alpha}\right)=0$
. Since this is trivially satisfied it will hold also after
coupling to other fields. In particular, in \cite{jpa2}, we have
coupled the self-dual model to charged scalar fields by using an
arbitrary constant ``$a$'' as follows: $\p_{\mu}\phi^*\p^{\mu}\phi
\to (D_{\mu}\phi)^*D^{\mu}\phi + e^2(a-1)f^2 \phi^*\phi $, where
``$e$'' is the charge and $D_{\mu}\phi = \left(\p_{\mu} + i\, e
f_{\mu}\right)\phi$. We have shown in \cite{jpa2} that the Bianchi
identity $\p_{\mu}F^{\mu}(A)=0$ gives rise via dual map to the
constraint $\p_{\mu} \left\lbrace \left\lbrack m^2 + 2 \, e^2
(a-1)\phi^* \phi \right\rbrack f^{\mu}\right\rbrace =0$. Although
only for $a=1$ we have a ``minimal coupling'', the correct
counting of degrees of freedom is guaranteed for any value of
``$a$''. In the spin two case the traceless condition $f=0$ does
not correspond to a trivial identity in the dual gauge theories.
Therefore we expect restrictions on the possible couplings of the
spin two self-dual model to other fields.

Concerning the local symmetries of the models $S_{SD}^{(2)}$ and
$S_{SD}^{(3)}$ a comment is in order. Namely, the first term in
$S_{SD}^{(1)}$ is invariant under the local transformations
$\delta_{\xi} \fab = \p_{\alpha}\xi_{\beta} $. This symmetry is
broken by the Fierz-Pauli mass term. However, in the dual theory
$S_{SD}^{(2)}$ such symmetry is restored. Analogously, the first
term in $S_{SD}^{(2)}$ is invariant under antisymmetric local
shifts $\delta_{\Lambda} A_{\alpha\beta} = \Lambda_{\alpha\beta}$,
where $\Lambda_{\alpha\beta}=-\Lambda_{\beta\alpha}$, and that
symmetry is broken by the mass term of $S_{SD}^{(2)}$ (CS$_1$
term). Once again the symmetry is restored in $S_{SD}^{(3)}$ which
depends only on $h_{(\alpha\beta)}$. Since both the quadratic
Einstein-Hilbert action and the mass term (quadratic third order
Chern-Simons term) of $S_{SD}^{(3)}$ are invariant under the same
set of local symmetries there will be no local symmetry to be
restored by a higher (fourth) order self-dual model. So we claim
that $S_{SD}^{(3)}$ is the highest order spin two self-dual model.
Likewise, in the spin one case both the Maxwell and Chern-Simons
terms are invariant under the same gauge symmetry and we have no
third order self-dual model of spin one.

\section{Generalized self-dual model of spin two and its dual}

In the last section we have learned that there are at least three
different consistent ways of giving mass to a parity singlet of
spin two in $D=2+1$ without using extra fields. We can use the
Fierz-Pauli mass term, the CS$_1$ term or the Chern-Simons term of
third order which is a quadratic truncation of a gravitational
Chern-Simons term, see (\ref{lsd1}), (\ref{sd2}) and (\ref{sd3})
respectively. In the spin one case (parity singlet) we have two
possible mass terms, i.e., the first order Chern-Simons term and
the Proca term which appears in the first order self-dual model of
\cite{tpn}. Both terms can coexist in a generalized self-dual
model (Maxwell-Chern-Simons-Proca theory) which contains two
massive parity singlets of spin one in the spectrum. It is
natural\footnote{In a more general situation we might try to
combine the three different spin two mass terms altogether
\cite{ddelm}} to ask whether we could combine different mass terms
also in the spin two case. Indeed, this question has been
addressed  in \cite{tekin}. As we have seen here in passing from
$S_{SD}^{(1)}$ to $S_{SD}^{(3)}$ the sign of the Einstein-Hilbert
term changes, which poses a problem when both Fierz-Pauli and the
topological Chern-Simons term (quadratic truncation) of
$S_{SD}^{(3)}$ are present since they require opposite signs for
the Einstein-Hilbert action. In fact, due to this problem the
authors of \cite{tekin} have concluded that the theory consisting
of an Einstein-Hilbert action plus a topological Chern-Simons term
of third order and a Fierz-Pauli mass term does not have a
physical spectrum. On the other hand, we have seen that the sign
of the EH term in $S_{SD}^{(1)}$ and $S_{SD}^{(2)}$ is the same,
so it is expected that we could define a theory with two massive
physical particles in the spectrum by combining both mass terms of
$S_{SD}^{(1)}$ and $S_{SD}^{(2)}$. In analogy with  the spin one
case \cite{jhep2} we define a generalized self-dual model of spin
two by adding a quadratic Einstein-Hilbert term to the
$S_{SD}^{(1)}$ self-dual model defined with arbitrary coefficients
$a_0,a_1$:

\be S_{GSD} =  \int \left\lbrack \frac {a_0}{2} \left(f^2
\right)_{FP}  +  \frac {a_1}{2} f \cdot d\, f + \frac{f \cdot d\,
\Omega (f)}4 \right\rbrack \label{gsd} \ee

\no We could ask what is the gauge theory dual do $S_{GSD}$ which
generalizes $S_{SD}^{(2)}$. Following \cite{jhep2}, in order to
avoid ghosts, it is appropriate to introduce auxiliary fields
($\lambda_{\alpha\beta}$) and rewrite the quadratic EH term of
(\ref{gsd}) in a first order form with the help of a Fierz-Pauli
mass term. Next we add two terms, with no particle content, to mix
the initial fields ($f_{\alpha\beta},\lambda_{\alpha\beta}$) with
the new dual fields
($\tilde{A}_{\alpha\beta},\tilde{B}_{\alpha\beta}$). Introducing a
source term we have the generating function

 \be W\left\lbrack j
\right\rbrack \, = \, \int {\cal D}\tilde{A}\, {\cal D}\tilde{B}
\, {\cal D}f \, {\cal D}\lambda \, \exp \, i \, S_M(j)  \qquad ,
\label{wj} \ee

\no where the source dependent master action is given by

\bea  S_M(j)  &=&  \frac {a_0}{2} \int \left(f^2 \right)_{FP} +
\frac {a_1}{2} \int f \cdot d\, f + \int d^3 x \,
j^{\mu\nu}f_{\mu\nu}
\nn \\
&+& \frac {1}{2} \int \left(\lambda^2 \right)_{FP}  + \int \lambda
\cdot d\, f  \label{smj1}\\
&-& \int ( \lambda - \tilde{B})  \cdot d\, (f-\tilde{A}) - \frac
{a_1}{2} \int (f-\tilde{A}) \cdot d\, (f-\tilde{A})  \nn \eea

\no After the shifts $\tilde{B}_{\alpha\beta} \to
\tilde{B}_{\alpha\beta} + \lambda_{\alpha\beta}$ and
$\tilde{A}_{\alpha\beta} \to \tilde{A}_{\alpha\beta} +
f_{\alpha\beta}$ in $S_M$ the last two terms decouple and since
they have no propagating mode, the particle content of $S_M$ is
the same of the generalized self-dual model $S_{GSD}$. Integrating
over $\tilde{A} , \tilde{B} $ and $\lambda_{\alpha\beta}$ we
obtain the generating function of the GSD model up to an overall
constant:

\be W\left\lbrack j \right\rbrack =  \int {\cal D}f\,
e^{i\left\lbrack S_{GSD}(f) + \int d^3 x\, j^{\mu\nu} f_{\mu\nu}
\right\rbrack }\label{wjgsd} \ee

\no On the other hand we can write:

\bea S_M(j)  &=& - \int \tilde{B} \cdot d\, \tilde{A} - \frac
{a_1}{2} \int \tilde{A} \cdot d\, \tilde{A} + \int d^3 x \,
j^{\mu\nu} f_{\mu\nu} \nn \\
&+& \frac {1}{2} \int \left(\lambda^2 \right)_{FP} +
 \int \lambda  \cdot d\, \tilde{A} \label{sm4} \\
&+& \frac {a_0}{2} \int \left(f^2 \right)_{FP} + \int f \cdot d
\left( \tilde{B} + a_1 \tilde{A}\right) \nn \eea

\no The integrals $\int {\cal D}\lambda$ and $\int {\cal D} f$
will produce two Einstein-Hilbert terms quadratic in the fields
$\tilde{A}_{\alpha\beta}$ and $\tilde{B}_{\alpha\beta}$ including
a mixing term involving both fields. A field redefinition can
decouple $\tilde{A}_{\alpha\beta}$ from $\tilde{B}_{\alpha\beta}$
. Guided by the spin one case \cite{jhep2} we use the convenient
notation:

\be a_0 = m_+ m_- \, ; \, a_1 = m_+ - m_-  \label{a0} \ee

\no After the redefinitions:

\bea \tilde{A}_{\alpha\beta} &=& \frac{\sqrt{m_+} A_{\alpha\beta}
- \sqrt{m_-}
B_{\alpha\beta}}{\sqrt{m_+ + m_-}} \label{t1} \\
\tilde{B}_{\alpha\beta} &=& - \frac{m_+^{3/2} A_{\alpha\beta} +
m_-^{3/2} B_{\alpha\beta}}{\sqrt{m_+ + m_-}} \label{t2} \eea

\no we deduce, up to an overall constant,

\be W\left\lbrack j \right\rbrack = \int {\cal D}A \, {\cal D}B \,
e^{i \, S\left\lbrack j,m_+,m_-\right\rbrack } \label{wjs} \ee

\no where

\bea S \left\lbrack j,m_+,m_-\right\rbrack &=& S_{SD}^{(2)}(A,m_+)
+
S_{SD}^{(2)}(B,-m_-) \nn \\
&+& \int d^3 x\, \left\lbrack j^{\alpha\nu}F_{\alpha\nu}(A,B) +
\frac{j^{\alpha\nu}j_{\nu\alpha}}{2 m_+ m_-} - \frac{j_{\mu}^{\mu}
j_{\alpha}^{\alpha}}{4 m_+ m_-} \right\rbrack \label{spd} \eea

\no The tensor $F_{\alpha\nu}(A,B)$ is invariant under independent
gauge transformations $\delta A_{\alpha\beta} =
\p_{\alpha}\xi_{\beta} $ and $\delta B_{\alpha\beta} =
\p_{\alpha}\zeta_{\beta} $, explicitly:

\bea F_{\alpha\nu} (A,B) &=&
\epsilon_{\alpha\beta\gamma}\p^{\beta}C^{\gamma}\, _{\nu} -
\frac{\eta_{\alpha\nu}}2
\epsilon^{\mu\gamma\lambda}\p_{\mu}C_{\gamma\lambda} \label{fab3} \\
 C_{\alpha\beta} &=& - \frac{1}{\sqrt{m_+ + m_-}}\left(\frac{A_{\alpha\beta}}{\sqrt{m_+}} +
  \frac{B_{\alpha\beta}}{\sqrt{m_-}}\right) \label{cab} \eea

\no  For $m_+ = m_-$ parity symmetry is restored in both
(\ref{gsd}), using (\ref{a0}), and (\ref{spd}). Using the physical
interpretation of $S_{SD}^{(2)}$ from the last section it is now
clear that $S_{GSD}$ describes two massive particles of masses
$m_+$ and $m_-$ and helicities $+2$ and $-2$. Comparing
correlation functions from (\ref{wjgsd}) and (\ref{wjs}) we derive
:

\bea \left\langle  f_{\mu_1\nu_1}(x_1)\cdots
f_{\mu_N\nu_N}(x_N)\right\rangle_{S_{GSD}(f,m_+,m_-)} &=&
\left\langle F_{\mu_1\nu_1}\left\lbrack C(x_1)\right\rbrack \cdots
F_{\mu_N\nu_N}\left\lbrack C(x_N) \right\rbrack
\right\rangle_{S_{SD}^{(2)}(A,m_+) + S_{SD}^{(2)} (B,-m_-)} \nn\\
&+& \, {\rm contact \, \, terms}  \label{cf3}\eea

\no So we have the map $\fab \leftrightarrow F_{\alpha\beta}(C)$.
For a complete proof of equivalence between $S_{GSD}(f,m_+,m_-)$
and the gauge invariant sector of $S_{SD}^{(2)}(A,m_+) +
S_{SD}^{(2)}(B,-m_-)$ it is rather puzzling that $\fab$ is mapped
into a gauge invariant function of one specific linear combination
of the fields $A_{\alpha\beta}$ and $B_{\alpha\beta}$ while on the
other side we have two independent and local gauge invariant
objects namely, $T_{\mu\alpha}(A)=\epsilon_{\mu}\,
^{\nu\lambda}\p_{\nu}A_{\lambda\alpha}/m $ and $T_{\mu\alpha}(B)$.
We should be able to compute any correlation function of
$T_{\mu\alpha}(A)$ and $T_{\mu\alpha}(B)$ in terms of the
generalized self-dual field $\fab$. Indeed, as in the spin one
case \cite{jhep2}, this is possible as we next show. We first
suppress the source term $\fab j^{\alpha\beta}$ in (\ref{smj1})
and add sources for $T_{\mu\alpha}(A)$ and $T_{\mu\alpha}(B)$. So
we define the generating function

\be \tilde{W}\left\lbrack \tj_+ , \tj_- \right\rbrack =  \int
{\cal D}f \, {\cal D}\lambda \, {\cal D}\tilde{A}\, {\cal
D}\tilde{B} \, \exp \, i \tilde{S}_M\left\lbrack \tj_+ , \tj_-
\right\rbrack \label{wt} \ee

\no where

\be \tilde{S}_M \left\lbrack \tj_+ , \tj_- \right\rbrack = S_M
(j=0) + \int \, d^3 x \, \left\lbrack \tj_+^{\mu\alpha}
T_{\mu\alpha}(\tilde{A}) + \tj_-^{\mu\alpha}
T_{\mu\alpha}(\tilde{B}) \right\rbrack \qquad . \label{st} \ee

\no We have introduced the sources

\bea &&\tj_+ \equiv \frac 1{\sqrt{m_+ + m_-}} \left(\frac{ m_-
j_+}{\sqrt{m_+}} - \frac{m_+
j_-}{\sqrt{m_-}}\right) \\
&&\tj_- \equiv - \frac 1{\sqrt{m_+ + m_-}} \left(
\frac{j_+}{\sqrt{m_+}} + \frac{j_-}{\sqrt{m_-}}\right) \eea

\no in a such way that after integration over $\fab$ and
$\lambda_{ab}$ and redefining the fields according to (\ref{t1})
and (\ref{t2}) we obtain, up to an overall constant,

\bea &&W\left\lbrack j_+ , j_- \right\rbrack
 = \tilde{W}\left\lbrack \tj_+ , \tj_- \right\rbrack = \nn\\ &&\int
{\cal D}f \, {\cal D}\lambda \, {\cal D}A\, {\cal D}B \exp \, i
\left\lbrace S_{SD}^{(2)}(A,m_+) + S_{SD}^{(2)}(B,-m_-) + \int \,
d^3 x \, \left\lbrack j_+^{\mu\alpha} T_{\mu\alpha}(A) +
j_-^{\mu\alpha} T_{\mu\alpha}(B)\right\rbrack \right\rbrace \nn\\
\label{w2} \eea

\no On the other hand, it is not difficult to convince oneself
that after some shifts of $\tilde{B}_{\alpha\beta}$ and
$\tilde{A}_{\alpha\beta}$ in
 (\ref{wt}) we can decouple those fields completely.
Their integration leads to a constant. By further integrating over
the auxiliary fields $\lambda_{\alpha\beta}$ we obtain from
(\ref{wt}), up to an overall constant, the dual version of
(\ref{w2}),

\bea && W\left\lbrack j_+ , j_- \right\rbrack
 = \tilde{W}\left\lbrack \tj_+ , \tj_- \right\rbrack = \nn\\ &&\int
{\cal D}f \,  \exp \, i \left\lbrace S_{GSD}(f) +  \int \, d^3 x
\, \left\lbrack j_+^{\lambda\alpha}
D_{\lambda\alpha}\,^{\mu\nu}(x,-m_-)f_{\mu\nu} +
j_-^{\lambda\alpha}D_{\lambda\alpha}\,^{\mu\nu}(x,m_+)f_{\mu\nu}
\right\rbrack  + {\cal O}(j^2) \right\rbrace \nn \\ \label{w2dual}
\eea

\no where ${\cal O}(j^2)$ stand for quadratic terms in the sources
$j_+$ and $j_-$. We have introduced the differential operator:

\be D^{\lambda\alpha\mu\nu} (x,m) = \frac{1}{\vert m \vert
\sqrt{m_+ + m_-}}\left\lbrack m \,
E^{\lambda\mu}_x\eta^{\alpha\nu}- E_x^{\lambda (\mu} E_x^{\nu )
\alpha}\right\rbrack  \label{d} \ee

\no Note that (\ref{w2}) and (\ref{w2dual}) are both symmetric
under $(m_+,m_-,j_+,j_-) \to (-m_-,-m_+,j_-,j_+)$  as expected.
Correlation functions of $T_{\mu\alpha}(A)$ and $T_{\mu\alpha}(B)$
can now be calculated from the GSD model. For instance, from
(\ref{w2}) and (\ref{w2dual}) we derive:

\bea && \left\langle T^{\alpha_1\beta_1}\left\lbrack
A(x_1)\right\rbrack \cdots T^{\alpha_N\beta_N}\left\lbrack
A(x_N)\right\rbrack \right\rangle_{S_{SD}^{(2)}(A,m_+) + S_{SD}^{(2)}(B,-m_-)} = \nn \\
&& D^{\alpha_1\beta_1\mu_1\nu_1}(x_1,m_+) \cdots
D^{\alpha_N\beta_N\mu_N\nu_N}(x_N,m_+) \left\langle
f_{\mu_1\nu_1}(x_1) \cdots f_{\mu_N\nu_N}(x_N)
\right\rangle_{S_{GSD}} + {\rm contact \, terms}\nn \\
\label{cf4} \eea

\no Of course, we can also calculate correlation functions of
$T_{\mu\alpha}(B)$ and  mixed correlation functions involving both
$T_{\mu\alpha}(A)$ and $T_{\mu\alpha}(B)$ from the GSD model
(\ref{gsd}). So we prove the quantum equivalence between the gauge
invariant sector of $S_{SD}^{(2)}(A,m_+) + S_{SD}^{(2)}(B,-m_-)$
and the GSD model, up to contact terms. The classical equivalence
between those models can also be established in a analogous
fashion to what has been done in the spin one case in
\cite{jhep2}.

\section{Conclusion}

\no We have shown in the master action approach how duality can
help us to prove the decoupling of redundant degrees of freedom at
quantum level. We have  compared correlation functions and derived
a dual map between the first, second and third order self-dual
models which describe parity  singlets of helicity $+2$ (or $-2$)
in $D=2+1$. In particular, the decoupling of the antisymmetric
combinations $f_{\lbrack \alpha\beta \rbrack }$ and the transverse
conditions
$\p_{\alpha}f^{\alpha\beta}=0=\p_{\beta}f^{\alpha\beta}$ have been
shown to be related via dual maps to the trivial (non-dynamical)
identities $T_{\alpha\beta}(\Omega)-T_{\beta\alpha}(\Omega)=0$ and
$\p_{\alpha}T^{\alpha\beta}(\Omega)=\p_{\alpha}\left(\epsilon^{\alpha\nu\gamma}\p_{\nu}\Omega_{\gamma}\,^{\beta}\right)=0$
respectively, which indicates that those constraints will be no
obstacles for the inclusion of interactions, contrary to the
traceless condition $f_{\mu}^{\mu}=0$. Furthermore, we have seen
that the spectrum equivalence of the three self-dual models
follows from the non-propagating (pure gauge) nature of the mixing
terms in the master action, namely, the Chern-Simons term of first
order and the Einstein-Hilbert action. Based on the local
symmetries of the self-dual models we have argued why we should
not expect a fourth or higher order self-dual model of spin two
and why there is no third (or higher) order  self-dual model in
the spin one case.

In section 3 we have defined a generalized self-dual model (GSD)
by adding a quadratic Einstein-Hilbert term to the first order
self-dual model of \cite{aragone} and shown its equivalence to the
gauge invariant sector of a couple of noninteracting free
particles of opposite helicities ($+2$ and $-2$) and different
masses, i.e., $S_{SD}^{(2)}(A,m_+) + S_{SD}^{(2)}(B,-m_-)$. This
generalizes previous works \cite{deser1,ilha,scaria}. We have
identified (dual map) the gauge invariant field of the GSD model
with a  gauge invariant function of one specific linear
combination of the opposite helicity gauge fields, see
(\ref{cab}). In the opposite direction we have also shown how to
compute correlation functions of gauge invariant objects of
$S_{SD}^{(2)}(A,m_+) + S_{SD}^{(2)}(B,-m_-)$ from the dual GSD
theory. No specific gauge condition has been used.

 The decoupling of spurious degrees of freedom after
the inclusion of interactions is under investigation. It is also
of interest to formulate consistent self-dual models for higher
spin ($s\ge 3$) massive particles in $D=2+1$ since the cases $s=1$
and $s=2$ seem to indicate, as we have seen here, a connection
between topological actions and self-dual models. Finally, since
there are dimensional reductions from massless particles in $D+1$
to massive particles in $D$ dimensions, one might wonder which
mechanisms or which dual massless spin two models in $D=4$ give
rise to the three self-dual models described here in a unified
way.

\section{Acknowledgements}

D.D. is partially supported by \textbf{CNPq} while E.L.M. is
supported by \textbf{FAPESP} (06/59563-0). We thank discussions
with Alvaro de Souza Dutra and Marcelo Hott.


\begin{thebibliography}{99}

\bibitem{vasiliev} X. Bekaert et. al., ``Nonlinear higher spin theories in various dimensions'',hep-th/0503128.

\bibitem{sandrine} S. Cnockaert, ``Higher spin gauge field theories'', PhD Thesis, hep-th/0606121.

\bibitem{zinoviev} Yu. M. Zinoviev, Nucl.Phys.B {\bf 770} 83 (2007).

\bibitem{gaitan}R. Gaitan, arXiv:0711.2498, ``On the Coupling Problem of Higher Spin Fields in 2+1 Dimension '',
 PhD thesis, in spanish.

\bibitem{montemayor}  A. Khoudeir, R. Montemayor, Luis F. Urrutia,  arXiv:0806.4558.

\bibitem{kumar}  R. Banerjee, S. Kumar and S. Mandal Phys. Rev. D \textbf{63}, (2001)
125008.

\bibitem{jhep2} D. Dalmazi, JHEP 0608:040,2006, hep-th/0608129.

\bibitem{tekin} S. Deser and B. Tekin, Class. Q. Grav. {\bf 19} L97 (2002).


\bibitem{djt} S. Deser, R. Jackiw and S. Templeton, Ann. of Phys. {\bf 140}(1982)
372.

\bibitem{aragone} C. Aragone and A. Khoudeir, Phys. Lett.
B{\bf173} 141 (1986).

\bibitem{tpn} P.K. Townsend, K. Pilch and P. van Nieuwenhuizen, Phys. Lett B {\bf 136} (1984)38.

\bibitem{fierz} M. Fierz, Helv. Phys. Acta {\bf 12} (1939) 3; M.
Fierz, W. Pauli, Proc. Roy. Soc. {\bf 173} (1939) 211.

\bibitem{deser} S. Deser and J. McCarthy, Phys. Lett. B{\bf 246}
441 (1990).

\bibitem{jhep1} D. Dalmazi, JHEP 0601:132,2006, hep-th/0510153.

\bibitem{dj} S.Deser and R. Jackiw, Phys.Lett.B {\bf 139} (1984)
371.

\bibitem{botta} M. Botta Cantcheff,  Phys.Lett.B {\bf 533} (2002) 126.

\bibitem{jpa2} D. Dalmazi and Elias L. Mendon\c ca, J.Phys. A {\bf
39} (2006) 9355.

\bibitem{ddelm} D. Dalmazi and Elias L. Mendon\c ca (in progress).

\bibitem{deser1} S. Deser, gr-qc/9211010.

\bibitem{ilha} A. Ilha and C. Wotzasek, Phys. Rev. D {\bf 63},
(2001) 105013.

\bibitem{scaria} T. Scaria, ``Studies in certain planar field theories'', PhD Thesis, hep-th/0407022.

\end{thebibliography}
\end{document}